\documentclass[preprint,12pt]{elsarticle}

\usepackage{amssymb}
\usepackage{amsmath}

\journal{Journal of Subatomic Particles and Cosmology}

\begin{document}

\begin{frontmatter}



\title{Can high-$p_\perp$ theory and data constrain $\eta/s$?} 


\author[UWr]{Bithika Karmakar} 
\affiliation[UWr]{organization={Incubator of Scientific Excellence—Centre for Simulations of Superdense Fluids,\\
University of Wroc{\l}aw},
            addressline={pl M Borna 9}, 
            city={Wroc{\l}aw},
            postcode={50-204}, 
            country={Poland}}

\author[IPB]{Dusan Zigic}
\affiliation[IPB]{organization={Institute of Physics Belgrade,
University of Belgrade},
city={Belgrade},
postcode={11080}, 
country={Serbia}}

\author[IPB]{Igor Salom}

\author[UJ]{Jussi Auvinen}
\affiliation[UJ]{organization={ Department of Physics, University of Jyvaskyla},
addressline={P.O. Box 35},
postcode={FI-40014},
country={ Finland}} 

\author[UWr]{Pasi Huovinen}

\author[UB]{Marko Djordjevic}
\affiliation[UB]{organization={ Faculty of Biology, University of Belgrade},
city={Belgrade},
postcode={11000}, 
country={Serbia}}

\author[IPB]{Magdalena Djordjevic}

\begin{abstract}
Understanding the temperature dependence of the specific shear viscosity $(\eta/s)$ is crucial for characterizing the properties of the QCD matter produced in ultrarelativistic heavy-ion collisions. Since, low-$p_\perp$ theory and data are only weakly sensitive to the typical forms of $\eta/s(T)$, especially at high temperatures, we use high-$p_\perp$ data and theory to impose additional constraints on it. Our approach, based on dynamical radiative and collisional energy loss of high-$p_\perp$ particles, provides promising results in constraining the temperature dependence of $\eta/s$.
\end{abstract}

\begin{graphicalabstract}
\end{graphicalabstract}

\begin{highlights}
\item Constraining the temperature dependence of the specific shear viscosity $(\eta/s)$ of the medium created in heavy-ion collisions jointly using low- and high-$p_\perp$ data and theory. 
\item 
In the phenomenological approach, jet energy loss, high-$p_\perp$ $R_{AA}$, and flow harmonics are computed within the DREENA framework for three different medium evolutions, each characterized by a different $\eta/s(T)$. The theoretical predictions are then compared with experimental data to constrain $\eta/s$.
\item 
In the theoretical approach, 
 $\eta/s$ is obtained from the jet transport coefficient, which is calculated using the dynamical jet energy loss formalism.
\end{highlights}


\end{frontmatter}



\section{Introduction}
\label{sec1}
Quark-Gluon Plasma (QGP) created in ultrarelativistic heavy-ion collisions is found to behave as a nearly inviscid fluid. Typically, the specific shear viscosity ($\eta/s$) of the medium is studied by comparing the hydrodynamical simulation predictions to the low-$p_\perp$ data. The temperature of the system changes significantly during the collision. As a result, even if the QGP behaves like a nearly perfect fluid near $T_c$ (the “soft”, strongly
coupled regime), its $\eta/s$ may increase at higher temperatures where the QGP becomes weakly
coupled (the “hard”, weakly coupled regime). We refer this possibility as
the “soft-to-hard” medium hypothesis. However, determining the temperature dependence of $\eta/s$ throughout the entire evolution is surprisingly challenging. This is because the value of shear viscosity at higher temperatures has only a weak effect on anisotropies in collisions at
LHC energies, and almost no effect at all at RHIC.  Recent Bayesian analyses have constrained $\eta/s$ reasonably well in the temperature range
$T_c< T < 1.5T_c$, but the constraints are much weaker at higher temperature~\cite{Bernhard:2019bmu,Auvinen:2020mpc}. 

On the other hand, high-$p_\perp$ particles can serve as a complementary probe to the bulk properties of QGP. They are produced in the early stages of the collisions, propagate through the medium, interact with the partons, and lose energy. Therefore, it is reasonable to expect that the high-$p_\perp$ observables are sensitive to the bulk properties of QGP. In this study, we use our high-$p_\perp$ energy loss formalism $`$DREENA'~\cite{Zigic:2021rku} to constrain $\eta/s$ of the medium created in ultrarelativistic heavy-ion collisions. 

\section{Formalism}
In the phenomenological approach, we investigate whether high-$p_\perp$ observables can constrain the temperature dependence of $\eta/s$ of QGP in $\sqrt{s}=5.02$ TeV Pb+Pb (LHC) and $\sqrt{s}=200$ GeV Au+Au (RHIC) collisions. We consider three different parametrizations of $\eta/s(T)$, as detailed below~\cite{Bernhard:2019bmu}.
\begin{eqnarray}
 (\eta/s)(T) = \left\{
   \begin{aligned}
      &(\eta/s)_{\text{min}}, &&  T<T_c, \\
      &(\eta/s)_{\text{min}} + (\eta/s)_{\text{slope}} (T-T_c)
                        \Big( \frac{T}{T_c} \Big)^{(\eta/s)_{\text{crv}}}, &&  T>T_c,
   \end{aligned}\right.
\end{eqnarray}
(i) constant $\eta/s$ (0.15 for Pb+Pb collision and 0.12
    for Au+Au collision ), (ii) $(\eta/s)_{\text{min}}$ = 0.1, $(\eta/s)_{\text{slope}}$= 1.11,
    $(\eta/s)_{\text{crv}}$ = -0.48, (we label it as $`$Nature') and (iii) $(\eta/s)_{\text{min}}$ = 0.04, $(\eta/s)_{\text{slope}}$ = 3.30,
    $(\eta/s)_{\text{crv}}$ = 0 (labelled as $`$LHHQ').

We generate $10^4$ events using T$_{\text{R}}$ENTo initial condition model, and sort them into centrality classes according to the number of participants. T$_{\text{R}}$ENTo normalization parameters are tuned to reproduce a subset of low-$p_\perp$ observables. The temperature evolution of the medium is obtained using the 2+1D hydrodynamical model $`$VISHNew'~\cite{Song:2007ux}. We then calculate jet energy loss and the resulting high-$p_\perp$ observables using the $`$generalized DREENA-A' framework, which takes the temperature profiles as input. These observables are compared with the experimental data to find out which $\eta/s(T)$ is preferred by the high-$p_\perp$ data.  

On the other hand, in the second approach, we compute $\eta/s$ directly from the high-$p_\perp$ energy loss formalism DREENA. DREENA (Dynamical Radiative and Elastic ENergy loss Approach) is based on the generalized Hard Thermal Loop (HTL) approach and includes both radiative and collisional energy loss in a dynamical, finite size QCD medium. It is generalized to include magnetic mass and running coupling constant. The transverse momentum loss of the jet in the elastic collisions with the medium partons is quantified by the jet quenching parameter $\hat q$. As shown in Ref.~\cite{MBW}, $T^3/\hat q$ is directly proportional to $\eta/s$ in a weakly coupled QCD medium. Therefore, in this theoretical approach, we compute $\eta/s(T)$ by evaluating $\hat q$ as a function of temperature.
In the fluid rest frame, the transport coefficient $\hat{q}$ can then be computed using the HTL resummed elastic collision rate as~\cite{JET:2013cls}
\begin{eqnarray}
\hat{q}  = \int_0^{\sqrt{6 E T}} d^2 q \, q^2 \cdot \frac{d \Gamma_{e l}}{d^2 q}.
\end{eqnarray}
Moreover, in the weakly coupled regime, $\eta/s$ can be related to $\hat q$ as~\cite{MBW, Muller}
\begin{eqnarray}
\eta/s \approx 1.25 \frac{T^3}{\hat q}.\label{etabys_qhat_reln}
\end{eqnarray}

\section{Results}
\begin{figure}[h]
\centering
\includegraphics[scale=0.7]{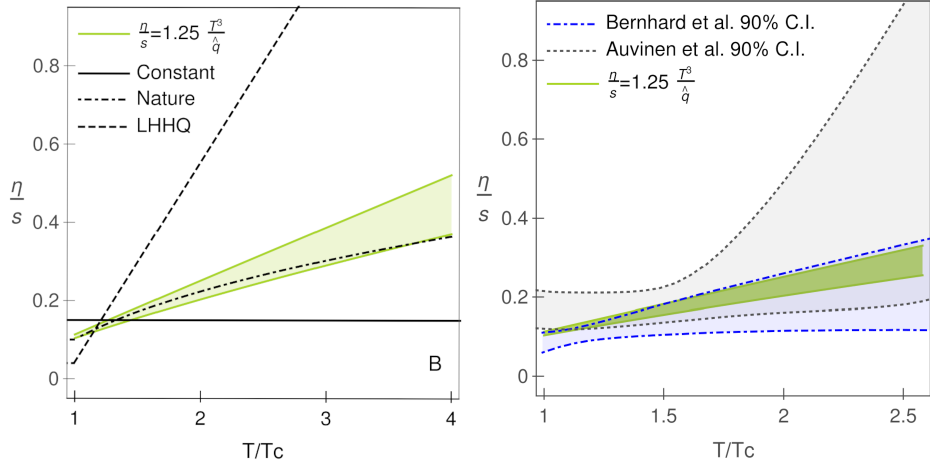}
\caption{$\eta/s$ computed from $\hat q$ is compared with the three $\eta/s(T)$ parametrizations (left panel). It is compared with the 90\% credible intervals of the Bayesian analyses
of Refs.~\cite{Bernhard:2019bmu,Auvinen:2020mpc} (right panel).}\label{figqhat}
\end{figure}

In Fig.~\ref{figRAA}, we compare the DREENA predictions for $R_{AA}$, high-$p_\perp$ flow harmonics $v_2$, $v_3$, and $v_4$ of charged hadrons, computed for three different $\eta/s(T)$ parameterizations, with the experimental data. Panels (a) and (b) show results for $\sqrt{s}=5.02$ TeV Pb+Pb collisions at 20-30\% and 40-50\% centralities, while panels (c) and (d) show the same for $\sqrt{s}=200$ GeV Au+Au collisions. As seen in the figure that the three cases are almost indistinguishable from each other, as the difference in temperature evolution in the three cases are too little to produce any visible differences in high-$p_\perp$ observables.
\begin{figure}[h]
\centering
\includegraphics[scale=0.2]{ch_PbPb_AuAu_label.pdf}
\caption{DREENA predictions for $R_{AA}$ and high-$p_\perp$ flow harmonics $v_2$, $v_3$, and $v_4$ of charged hadrons at 20-30\% and 40-50\% in Pb+Pb ($\sqrt{s}=5.02$ TeV) (panels a, b) and  Au+Au ($\sqrt{s}=200$ GeV) (panels c, d) collisions are shown and compared with experimental data. For more details, see Ref.~\cite{Karmakar:2023ity}.}\label{figRAA}
\end{figure}
The $\eta/s$ computed from the jet transport coefficient $\hat q$ via Eq.~\eqref{etabys_qhat_reln} is shown as a green band in the left panel of Fig.~\ref{figqhat}, along with the three $\eta/s(T)$ parametrizations considered in the study. Notably, the result inferred from our energy loss formalism agrees surprisingly well with the Bayesian analysis result ($`$Nature'). In the right panel, we compare our result~\cite{Karmakar:2023ity} with the Bayesian analysis results from Refs.~\cite{Bernhard:2019bmu,Auvinen:2020mpc}. It is evident that the uncertainty due to the jet initial energy is much smaller than that in the Bayesian analyses at high temperatures. Interestingly, our result does not drop below the Bayesian result beyond weakly coupled regime i.e., near $T_c$, despite the fact that our jet energy loss formalism is valid in the weakly coupled regime.

\section{Summary}

We use both the low- and high-$p_\perp$ sectors to constrain the temperature dependence of $\eta/s$ of the medium created in ultrarelativistic heavy-ion collisions. High-$p_\perp$ observables are calculated using DREENA for three different $\eta/s(T)$ parametrizations considered in this study. However, the temperature differences among the three evolutions are insufficient to lead to any visible differences in the high-$p_\perp$ observables. In contrast, in the second approach, we compute $\eta/s$ from the jet transport coefficient $\hat q$. The inferred result agrees very well with the low-$p_\perp$ Bayesian analysis result and gives way smaller uncertainty at high temperatures, where $\eta/s$ is otherwise hard to constrain. Since this agreement extends down to $T_c$, the soft-to-hard boundary can not be clearly identified.

{\em Acknowledgments:}
This work is supported by the European Research Council, grant
ERC-2016-COG: 725741 and Excellence Initiative–Research University of the University of Wroclaw of the Ministry of Education and Science.


\begin{thebibliography}{00}

\bibitem{Bernhard:2019bmu}
  J.~E.~Bernhard, J.~S.~Moreland and S.~A.~Bass,
  Nature Phys. \textbf{15}, no.11, 1113-1117 (2019).

\bibitem{Auvinen:2020mpc}
  J.~Auvinen, K.~J.~Eskola, P.~Huovinen, H.~Niemi, R.~Paatelainen and
  P.~Petreczky,
  Phys. Rev. C \textbf{102}, 044911 (2020).

\bibitem{Zigic:2021rku}
D.~Zigic, I.~Salom, J.~Auvinen, P.~Huovinen and M.~Djordjevic,
Front. in Phys. \textbf{10}, 957019 (2022).
\bibitem{Song:2007ux}
  H.~Song and U.~W.~Heinz,
  Phys.\ Rev.\ C \textbf{77}, 064901 (2008).


\bibitem{MBW}
  A.~Majumder, B.~Müller, and X.-N.~Wang,
  Phys.\ Rev.\ Lett.\ {\bf 99}, 192301 (2007).
  
\bibitem{JET:2013cls}
K.~M.~Burke \textit{et al.} [JET],
Phys. Rev. C \textbf{90}, 014909 (2014).

  \bibitem{Muller}
B.~M\"uller, Phys.\ Rev.\ D \textbf{104}, L071501 (2021).

\bibitem{Karmakar:2023ity}
B.~Karmakar, D.~Zigic, I.~Salom, J.~Auvinen, P.~Huovinen, M.~Djordjevic and M.~Djordjevic,
Phys. Rev. C \textbf{108}, no.4, 044907 (2023).


\end{thebibliography}
\end{document}